\documentclass[11pt]{article}
\date{}
\begin{document}
\sloppy
\title{ A Very Naive Model of Hadron with Negative Quintessence}
\author{V. Majern\'{\i}k \\
Institute of Mathematics, Slovak
Academy of Sciences, \\ SK-81473 Bratislava, \v Stef\'anikova  49,
Slovak Republic\\and\\  J\'anos Selye University, Kom\'arno, Slovak Republic}

\maketitle
\begin{abstract}
A simple model of hadron which exhibits the quark confinement and
the asymptotic freedom is described. The hadron is modelled as a
sphere of radius equal to its Compton wavelength in which quarks
occur surrounded by space with the uniformly distributed {\it
negative} vacuum energy density characterized by $\lambda$. The
acceleration stemming from this vacuum energy causes the quark
confinement. However, in the center of hadron quarks behave as
practically free particles. From the requirement that the Compton
wavelength of a hadron should remain constant while the scale
factor varies with time we estimate the strength of the vacuum
energy density surrounding quarks. It follows that the inward
pressure of this vacuum energy is large enough to keep quarks
inside the hadron.
 \footnote{E-mail:fyziemar@savba.sk}
\end{abstract}
\section{Introduction}

The recent astronomical observations
\cite{PER} \cite{pe} \cite{BO} give increasing
support for the accelerating and flat universe which
consists of a mixture of a small part of the
baryonic matter about one third non-relativistic dark matter (DM) and two
thirds of a smooth component, called dark energy (DE).
In this short communication, we put forward the hypothesis that
quarks in hadrons are surrounded by negative DE whose
inward pressure causes its confinement. This dark energy we
modelled by
{\it negative} cosmological constant $\Lambda$.

In the literature,
DE is theoretically modelled by many ways, e.g. as
(i) a very small cosmological constant (e.g.\cite{4}) (ii) quintessence
(e.g.\cite{5})
(iii) Chaplygin gas (e.g.\cite{6}) (iv) tachyon field (e.g.\cite{7}
\cite{PA} \cite{PAD}) (v)
interacting quintessence (e.g.\cite{8}),
 quaternionic field (e.g.\cite{IK}),
etc.
It is unknown which of the said models will finally
emerges as the successful
one.

Physicists found by the late 1960s that hadrons were made of quarks
which remain firmly locked together. Yet the deep-inelastic scattering
experiments probed the mechanism of quarks confinement shown that the
quarks behave practically as free particles (partons). This fact called
as asymptotic freedom refers to the vanishing of the strong nuclear force
between quarks as the distance between them approaches zero
\cite{GR} \cite{GRO} \cite{POL}.

\section{The model}

We model a hadron as a sphere of radius $R_m$  in which quarks occur
surrounded by space of the uniformly distributed
vacuum energy density characterized by $\lambda$.
From the Friedmann equations, it follows that the total energy of a unit
mass
situated at a distance R from the origin is given by the equation
\cite{H}
$$\frac{1}{2}\left (\frac{dR}{dt}\right )^2-\frac{GM}{R}
-\frac{c^2R^2 \lambda}{6}=-\frac{kc^2}{2}.\quad \eqno(1)$$
The first term is the kinetic energy, the second term is the
gravitational potential energy due to the mass contained in a sphere of
radius $R$, the third term may be interpreted as the potential energy due
to that portion of the vacuum encompassed by the same sphere. The
concomitant equation of motion is obtained by differentiating
Eq.(1) with respect to R \cite{H}
$$\ddot{R}=-\frac{GM}{R^2}+\frac{R\lambda}{3}.\quad\eqno(2)$$

According to Eq.(2), the acceleration between quarks in the field of
the {\it negative} vacuum energy (negative quintessence) which is
uniform
distributed inside a hadron is
$$\ddot{R}=-\frac{Gm_q}{R^2}-\frac{R\lambda}{3},\quad\eqno(3)$$
where $m_q$ is the mass of quark. The average mass density
of all quarks in a hadron is
$$\rho=\sum_{i=1}^{3} m_q^{i}/R_{h}^3\qquad \sum_{i=1}^{3}m_q^i=M_h,$$
where $m_q^i,i=1,2,3$ are masses of quarks and $M_h$ is the mass of the
whole hadron..

Here, the problem arises what value should be inserted for $\lambda$.
To estimate
this value we used the following argument.
The scale factor $R(t)$ is a universal fundamental
quantity depending on $t$ which
for the given stress-energy tensor and cosmological constant
is determined by the Einstein equations.
All distances in the universe should change in accord with the time
dependence of $R(t)$.
It is generally supposed that
the Compton wavelength of particles (except of photons of the
relict radiation)
does not varies
with the scale factor (remains constant). Next, we point out that this fact
can be also explained if the mass distribution in a hadron
is equal to the negative vacuum energy density.
We start with the standard Einstein field equations (c=1)
$$R_{ij}-g_{ij}(1/2)R=8\pi G(T^{(m)}_{ij}+T^{(v)}_{ij}),$$
where $T_{ij}^{(m)}$ is the energy-momentum tensor and
$$T^{(v)}_{ij}=\Lambda=\left (\frac{\lambda}{8\pi G}\right ) g_{ij}$$
the cosmological term.
In a homogeneous and isotropic medium characterized by the
Friedmann-Robertson-Walker line element the Einstein equations with
matter density $\rho$
and non-zero cosmical term $\Lambda$ acquire the following forms (k=1)
$$3{\dot R^2\over R(t)^2}= 8\pi G(\rho +\Lambda),
 \qquad \lambda=8\pi G \Lambda.\quad \eqno(4)$$
The necessary condition for the constancy of hadron radius is
$\dot R=0$ which implies according to Eq.(4) $\rho=-\Lambda$.

Keeping in mind that $\lambda =8\pi G \Lambda$ and $\Lambda\approx
-\rho$, respectively,
we have finally
$$\ddot R=T_1+T_2=-\frac{Gm_q}{R^2}-\frac{8 \pi G
R\rho}{3}.\quad\eqno(5)$$
We see that the first and second
term $T_1$ and $T_2$ in Eq.4 represents the acceleration of quarks in their
 gravitation field
and the acceleration
due to negative cosmological constant, respectively.
If the mass distribution within the sphere with the Compton wavelength
of a
hadron (e.g. neutron) $R_h\approx 10^{-15} m$ is homogeneous then the mass
density
$\rho_n\approx 10^{-45}$. Inserting $\rho_n$ into Eq.(5) and neglecting
the gravitational interaction between quarks we obtain
$$\ddot R=-\frac{8 \pi G \rho
}{3}R.\quad\eqno(6)$$
This is the familiar equation of motion of a harmonic oscillator $$\ddot
R=-\kappa R$$
with $\kappa \approx 5.10^{34}$.
Taking for mass of a quark the value $\approx 10^{-27} kg$ then the
force acting on it at $R_h$ reaches the value $\approx 10^{7} N$ which
is approximately $10^5$ larger than the force acting on a unit mass
at the Earth surface. This means that quarks in a hadron
are bound very strong.
\section{Conclusions}
(i) Quarks occurring
in a medium consisting of negative vacuum energy might be confined
due tho inward pressure of this vacuum energy.\\
(ii) From the requirement of constancy of the Compton wavelength of a
hadron  the density of the vacuum energy within the sphere
surrounding quarks
may be estimated.\\
(iii) The inward force acting on quarks due to negative vacuum
is strong enough to confine them within hadron.\\
(iv) Quarks occurring around the center of the sphere of the negative
vacuum energy behave as practically free particles.\\
(v) Hadrons appear as islands of quarks surrounded by negative vacuum
energy in a sea of the positive vacuum energy given by the common
cosmological constant.\\
(vi) The energy balance of positive quark energy and negative vacuum
energy surrounding them equals to zero.\\
(vii) The positive energy of quarks manifests itself by the gravitation
field in the neighboring space.\\

Of course, there are nowadays sophisticated theories for explaining the
asymptotical freedom of quarks. In this communication we only sketched
the basic idea of a model of hadron which point out that
the local concentration of negative vacuum energy might play an important
role not only in astrophysics \cite{VLD} \cite{BE} but also in the elementary particle
physics.

\end{document}